\documentclass[prl,twocolumn,showpacs,twoside,floatfix,fleqn]{revtex4}

\usepackage{amsfonts}
\usepackage{amssymb}
\usepackage{amsmath}
\usepackage{graphicx}

\begin{document}
\title{Underdoped cuprates phenomenology in the 2D Hubbard model within COM(SCBA)}
\author{Adolfo \surname{Avella}}
\email[E-mail: ]{avella@sa.infn.it}
\author{Ferdinando \surname{Mancini}}
\email[E-mail: ]{mancini@sa.infn.it} \homepage[Group Homepage:
]{http://scs.sa.infn.it}
\affiliation{Dipartimento di Fisica ``E.R.
Caianiello'' - Unit\`{a} CNISM di Salerno \linebreak Universit\`{a}
degli Studi di Salerno, I-84081 Baronissi (SA), Italy}

\begin{abstract}
The two-dimensional Hubbard model is studied within the Composite
Operator Method (COM) with the residual self-energy computed in the
Self-Consistent Born Approximation (SCBA). COM describes interacting
electrons in terms of the new elementary excitations appearing in
the system owing to strong correlations; residual interactions among
these excitations are treated within the SCBA. The anomalous
features appearing in the spectral function $A(\mathbf{k},\omega)$,
the momentum distribution function $n(\mathbf{k})$ and the Fermi
surface are analyzed for various values of the filling (from
overdoped to underdoped region) in the intermediate coupling regime
at low temperatures. For low doping, in contrast with the ordinary
Fermi-liquid behavior of a weakly-correlated metal found at high
doping, we report the opening of a pseudogap and some
non-Fermi-liquid features as measured for cuprates superconductors.
In addition, we show the presence of kinks in the calculated
electronic dispersion in agreement with ARPES data.
\end{abstract}

\date{July 25, 2006}


\maketitle

One of the most intriguing challenges in modern condensed matter
theory is the description of the anomalous behaviors experimentally
observed in novel materials. By anomalous behaviors we mean those
not predicted by standard many-body theory, that is, behaviors in
contradiction with Fermi-liquid framework and diagrammatic
expansions. Underdoped cuprates superconductors display anomalous
features in almost all experimentally measurable physical properties
\cite{Timusk_99,Damascelli_03,Eschrig_06}. As a matter of fact, the
microscopic description of this class of materials is still an open
problem because many of the anomalous features remain unexplained
or, at least, controversially debated \cite{Lee_06,Tremblay_06}:
non-Fermi-liquid response, quantum criticality, pseudogap formation,
ill-defined Fermi surface, and kinks in electronic dispersion.

Since the very beginning \cite{Anderson_87}, the two-dimensional
Hubbard model \cite{Hubbard_63} has been recognized as the minimal
model capable to describe the $Cu-O_2$ planes of cuprates
superconductors. It certainly contains many of the key ingredients
by construction: strong electronic correlations, competition between
localization and itineracy, Mott physics, and low-energy spin
excitations. Unfortunately, although fundamental for benchmarking
and fine tuning analytical theories, numerical approaches
\cite{Bulut_02} can be of little help to solve the puzzle of
cuprates owing to their limited resolution in frequency and
momentum. On the other hand, there are not so many analytical
approaches capable to deal with the quite complex aspects of
underdoped cuprates phenomenology. Among others, the most promising
approaches available in the literature are the cellular dynamical
mean-field theory \cite{Kotliar_01a}, the dynamical cluster
approximation \cite{Hettler_98}, and the cluster perturbation theory
\cite{Senechal_00}. Anyway, it is worth noticing that all these
approaches cannot avoid relying on some numerical method in order to
close their self-consistency cycles.

In this manuscript, we show that the two-dimensional Hubbard model
within a completely analytical self-consistent approach, the
Composite Operator Method (COM) \cite{Mancini_04} with the residual
self-energy computed in the Self-Consistent Born Approximation
(SCBA) \cite{Bosse_78}, can describe many of the anomalous features
contributing to the experimentally observed underdoped cuprates
phenomenology. In particular, we show how Fermi arcs develop out of
a large Fermi surface, how pseudogap shows itself in the dispersion
and in the density of states, how non-Fermi liquid features become
apparent in the momentum distribution function, and how much
\emph{kinked} can get the dispersion on varying doping. The
manuscript is organized as follows: first, we recall the Hubbard
model and fix the notation; then, we present the Composite Operator
Method and its application to the system under analysis; finally, we
present results and comparisons with experiments and give
conclusions.

The two-dimensional Hubbard model reads as
\begin{equation}
H=\sum_\mathbf{ij}(-\mu\delta_\mathbf{ij}-4t\alpha_\mathbf{ij})
c^\dag(i)c(j)+U\sum_\mathbf{i}n_\uparrow(i)n_\downarrow(i)
\end{equation}
where
$c^\dag(i)=\left(c_\uparrow^\dag(i),\,c_\downarrow^\dag(i)\right)$
is the creation electronic operator in spinorial notation and
Heisenberg picture ($i=(\mathbf{i},t_i)$), $\mathbf{i}$ is a vector
of the Bravais lattice, $n_\sigma(i)=c_\sigma^\dag(i)c_\sigma(i)$ is
the spin-$\sigma$ electronic number operator, $\mu$ is the chemical
potential, $t$ is the hopping integral and the energy unit, $U$ is
the Coulomb on-site repulsion,
$c^\alpha(\mathbf{i})=\sum_\mathbf{j}\alpha_\mathbf{ij}c(\mathbf{j})$
and $\alpha_\mathbf{ij}$ is the projector on the nearest-neighbor
sites.

COM recipe uses three main ingredients \cite{Mancini_04}:
\emph{composite} operators, \emph{algebra} constraints, and
\emph{residual} interactions treatment. Composite operators are
products of electronic operators and describe the new elementary
excitations appearing in the system owing to strong correlations.
According to the system under analysis \cite{Mancini_04}, you have
to choose a set of composite operators as operatorial basis and
rewrite the electronic operators and the electronic Green's function
in terms of this basis. You should think of composite operators just
as a better point, with respect to electronic operators, where to
start your mean field approximation. Algebra constraints are
relations among correlation functions dictated by the non-canonical
operatorial algebra closed by the chosen operatorial basis
\cite{Mancini_04}. After choosing an operatorial basis, one way to
obtain algebra constraints is to check which correlation functions
of two elements of the basis (named correlators hereafter) vanish or
can be expressed in terms of other correlators according to the
operatorial algebra closed by the basis. Other ways to obtain
algebra constraints rely on the symmetries enjoined by the
Hamiltonian under study, the Ward-Takahashi identities, the
hydrodynamics, etc \cite{Mancini_04}. You should think of algebra
constraints as a way to restrict the Fock space on which the chosen
operatorial basis acts to the Fock space of physical electrons.
Algebra constraints are used to compute unknown correlation
functions appearing in the calculations. Residual interactions among
the elements of the chosen operatorial basis are described by the
residual self-energy, that is, the propagator of the residual term
of the current after this latter has been projected on the chosen
operatorial basis \cite{Mancini_04}. According to the physical
properties under analysis and the range of temperatures, dopings,
and interactions you wish to explore, you have to choose an
approximation to compute the residual self-energy. You should think
of residual self-energy as a measure in the frequency and momentum
space of how much well defined, as quasi-particles, are your
composite operators.

Following COM prescriptions \cite{Mancini_04}, we have chosen as
operatorial basis
$\psi^\dag(i)=\left(\xi^\dag(i),\eta^\dag(i)\right)$, with
$\eta(i)=n(i)c(i)$ and $\xi(i)=c(i)-\eta(i)$, guided by the
hierarchy of the equations of motion and by the exact solution of
the Hubbard Hamiltonian reduced to its interacting term. The
retarded Green's function $G(i,j)=\left\langle
R\left[\psi(i)\psi^{\dagger}(j)\right]\right\rangle$ has the
following expression in terms of the two-pole propagator
$G^0(\mathbf{k},\omega)$
\begin{equation}
G(\mathbf{k},\omega)=\left(IG^0(\mathbf{k},\omega)^{-1}-\Sigma(\mathbf{k},\omega)I^{-1}\right)^{-1}I
\end{equation}
where $\Sigma=\mathcal{F}\left\langle R\left[\delta J(i)\delta
J^{\dagger}(j)\right]\right\rangle_I$ stands for the irreducible
part of the residual self-energy and $\mathcal{F}$ for the Fourier
transform. Till further notice, all objects appearing in the
equations stands for two by two matrices, according to the vectorial
nature of the operatorial basis. The entries of the normalization
matrix $I=\left\langle
\left\{\psi(\mathbf{i},t),\psi^{\dagger}(\mathbf{j},t)\right\}\right\rangle$
read as: $I_{11}=1-I_{22}$, $I_{22}=n/2$ and $I_{12}=I_{21}=0$. $n$
is the filling. The two-pole propagator $G^0(\mathbf{k},\omega)$ has
the following expression
\begin{equation}
G^0(\mathbf{k},\omega)=\left(\omega -
\varepsilon(\mathbf{k})\right)^{-1}I
\end{equation}
where $\varepsilon(\mathbf{k})$ is the energy matrix appearing in
the projected equations of motion of $\psi(i)$
\begin{equation}
\mathrm{i}\frac{\partial}{\partial
t}\psi(\mathbf{k},t)=\left[\psi(\mathbf{k},t),H\right]=\varepsilon(\mathbf{k})\psi(\mathbf{k},t)+\delta
J(\mathbf{k},t)
\end{equation}
once the constraint $\left\langle \left\{\delta
J(\mathbf{i},t),\psi^{\dagger}(\mathbf{j},t)\right\}\right\rangle=0$
has been enforced. This constraint assures that the residual current
$\delta J(i)$ describe the physics \emph{orthogonal} to the chosen
operatorial basis $\psi(i)$; that is, $\delta J(i)$ describes the
interactions among the elements of the operatorial basis.

Three parameters appear in the energy matrix
$\varepsilon(\mathbf{k})$: the chemical potential $\mu$, the
difference between upper and lower intra-subband contributions to
kinetic energy $\Delta=\left\langle \xi^{\alpha}(i)\xi^{\dagger }(i)
\right\rangle -\left\langle
\eta^{\alpha}(i)\eta^{\dagger}(i)\right\rangle$, and a combination
of the nearest-neighbor charge-charge, spin-spin and pair-pair
correlation functions $p=\frac{1}{4}\left\langle \delta
n_{\mu}^{\alpha}(i) \delta n_{\mu}(i)\right\rangle -\left\langle
\left[ c_{\uparrow}(i) c_{\downarrow}(i) \right]
^{\alpha}c_{\downarrow}^{\dagger}(i) c_{\uparrow}^{\dagger}(i)
\right\rangle$. $\delta n_{\mu}(i)= n_{\mu}(i)- \langle n_{\mu}(i)
\rangle$ stands for charge ($\mu=0$) and spin ($\mu=1,\,2,\,3$)
number operators and the sum over repeated indices is understood. By
exploiting algebra constraints and connections between propagators
and correlators, we have fixed the parameters appearing in the
energy matrix through a set of three self-consistent equations. Two
equations are obtained by expressing the filling $n$ and the
parameter $\Delta$ in terms of correlators, respectively. The third
equation is the algebra constraint $\left\langle \xi(i)\eta^{\dagger
}(i) \right\rangle=0$ that excludes double occupancy of a site by
two electrons with the same spin.

We have chosen to compute the residual self-energy
$\Sigma(\mathbf{k},\omega)$ within SCBA
\cite{Avella_03c,Krivenko_04,Mancini_04}. According to this, we have
\begin{equation}
\Sigma(\mathbf{k},\omega) = 4 t^2 I^{-1} S(\mathbf{ k},\omega)
\left(1-\sigma_x\right) I^{-1}
\end{equation}
with
$S(\mathbf{k},\omega)=\mathcal{F}_{\mathbf{k}}\left[S(\mathbf{r},\omega)\right]$
and
\begin{equation}
S(\mathbf{r},\omega)=\iint \frac{d\omega'd\Omega}{(2\pi)^2}
\frac{1+e^{-\beta\omega'}}{\omega-\omega'+\mathrm{i}\varepsilon}
F(\mathbf{r},\Omega)B(\mathbf{r},\omega'-\Omega)
\end{equation}
where
\begin{equation}
F(\mathbf{i-j},\omega)=\mathcal{F}_\omega\left\langle c^\alpha(i)
c^{\dag\alpha}(j)\right\rangle
\end{equation}
and
\begin{equation}\label{bosons}
B(\mathbf{i-j},\omega)=\mathcal{F}_\omega\left\langle \delta
n_{\mu}(i) \delta n_{\mu}(j) \right\rangle
\end{equation}
and $\mathcal{F}_\omega$ and $\mathcal{F}_{\mathbf{k}}$ are the
time-frequency and position-momentum Fourier transform operators,
respectively. $\sigma_x$ is the first Pauli matrix.

We have decided to compute both charge-charge and spin-spin
propagators (\ref{bosons}) within the two-pole approximation
\cite{Avella_03,Mancini_04} instead of using model spin
susceptibilities \cite{Plakida_06}. We have chosen charge and spin
number operators $n_{\mu}(i)$ and their currents $\rho_{\mu}(i) =
c^{\dagger}(i)\sigma_{\mu}c^{\alpha}(i)-c^{\dagger\alpha}(i)
\sigma_{\mu}c(i)$ as operatorial basis.  $\sigma_{\mu}=\left(
1,\,\mathbf{\sigma}\right)$ and $\mathbf{\sigma}$ are the Pauli
matrices. Within this framework, the bosonic propagators depend on
both electronic correlators and high-order bosonic correlation
functions, one per each channel, named $a_c$ and $a_s$
\cite{Avella_03,Mancini_04}. We have fixed $a_c$ and $a_s$ through
the algebra constraints $\left\langle n(i)n(i)\right\rangle=n+2D$
and $\left\langle n_z(i) n_z(i)\right\rangle=n-2D$, where $D$ is the
double occupancy, that excludes double occupancy of a site by two
electrons with the same spin and enforces the relation between
filling and \emph{lenght} of the electronic spin on the same site,
respectively.

\begin{figure}[tb]
\includegraphics[width=\columnwidth]{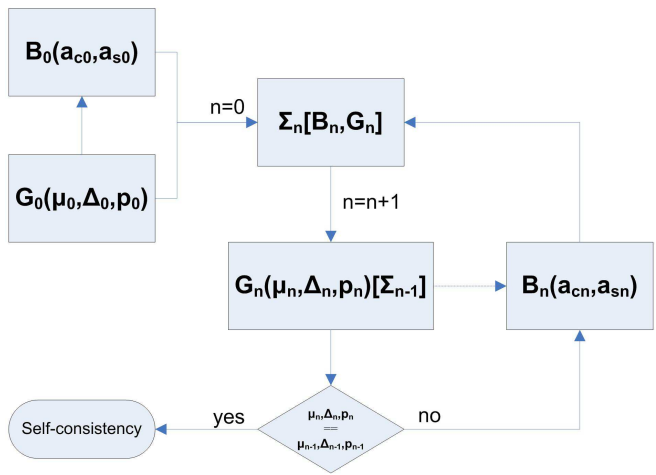}
\caption{Self-consistency scheme to compute the propagator $G$ in
terms of the charge-charge and spin-spin propagator $B$ and the
residual self-energy $\Sigma$.}\label{fig1}
\end{figure}

\begin{figure}[t!]
\includegraphics[width=\columnwidth]{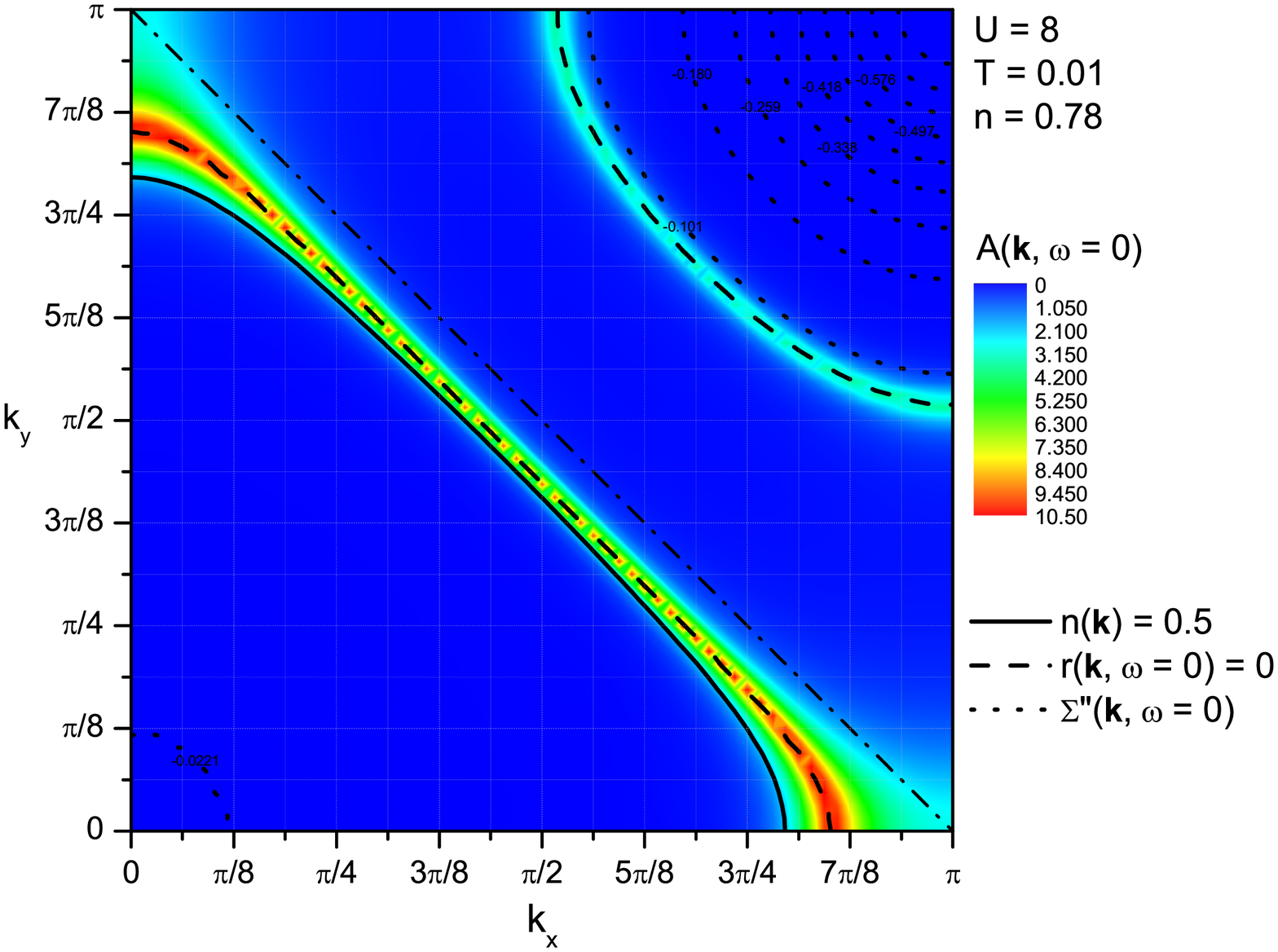}
\includegraphics[width=\columnwidth]{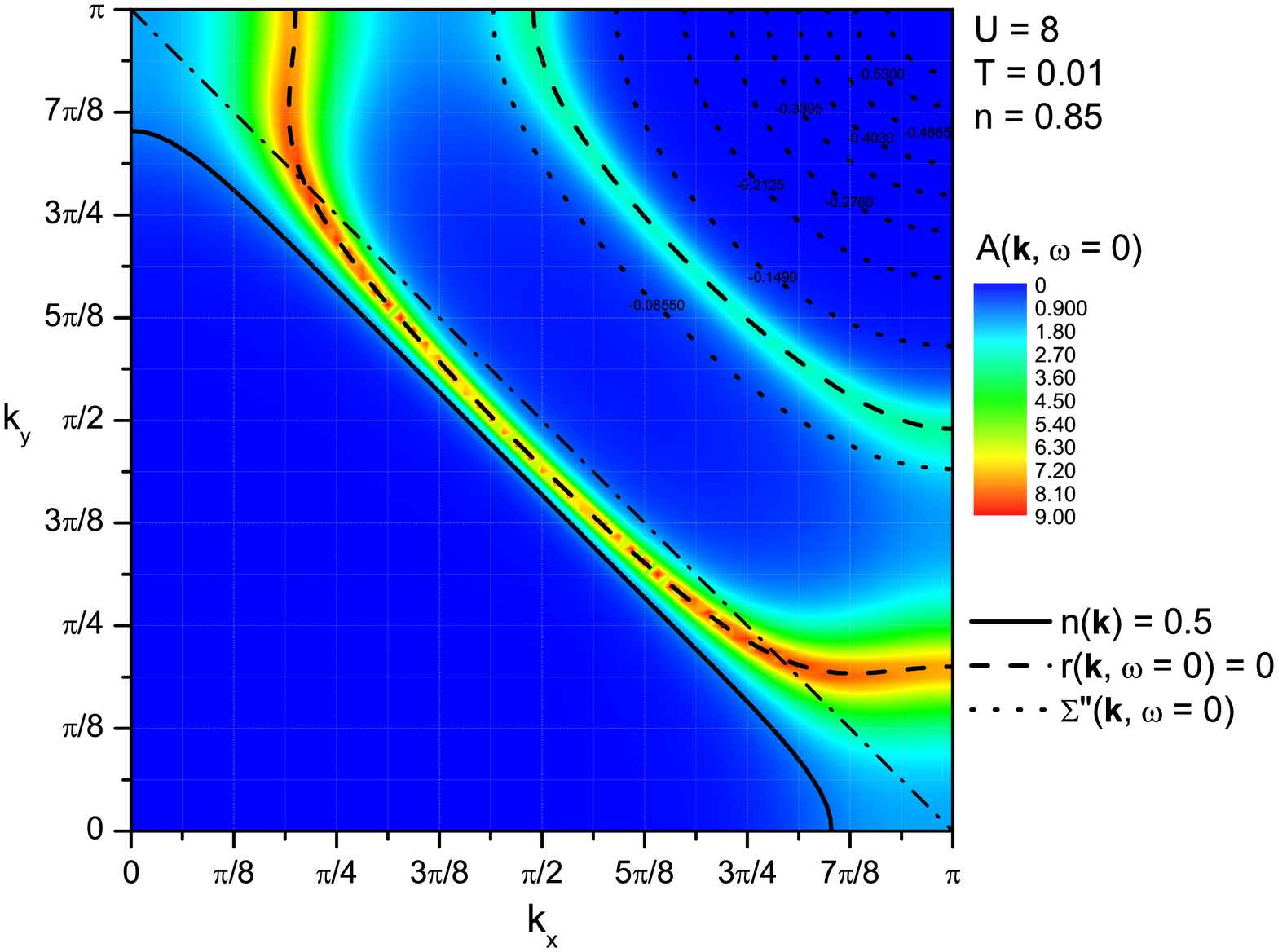}
\includegraphics[width=\columnwidth]{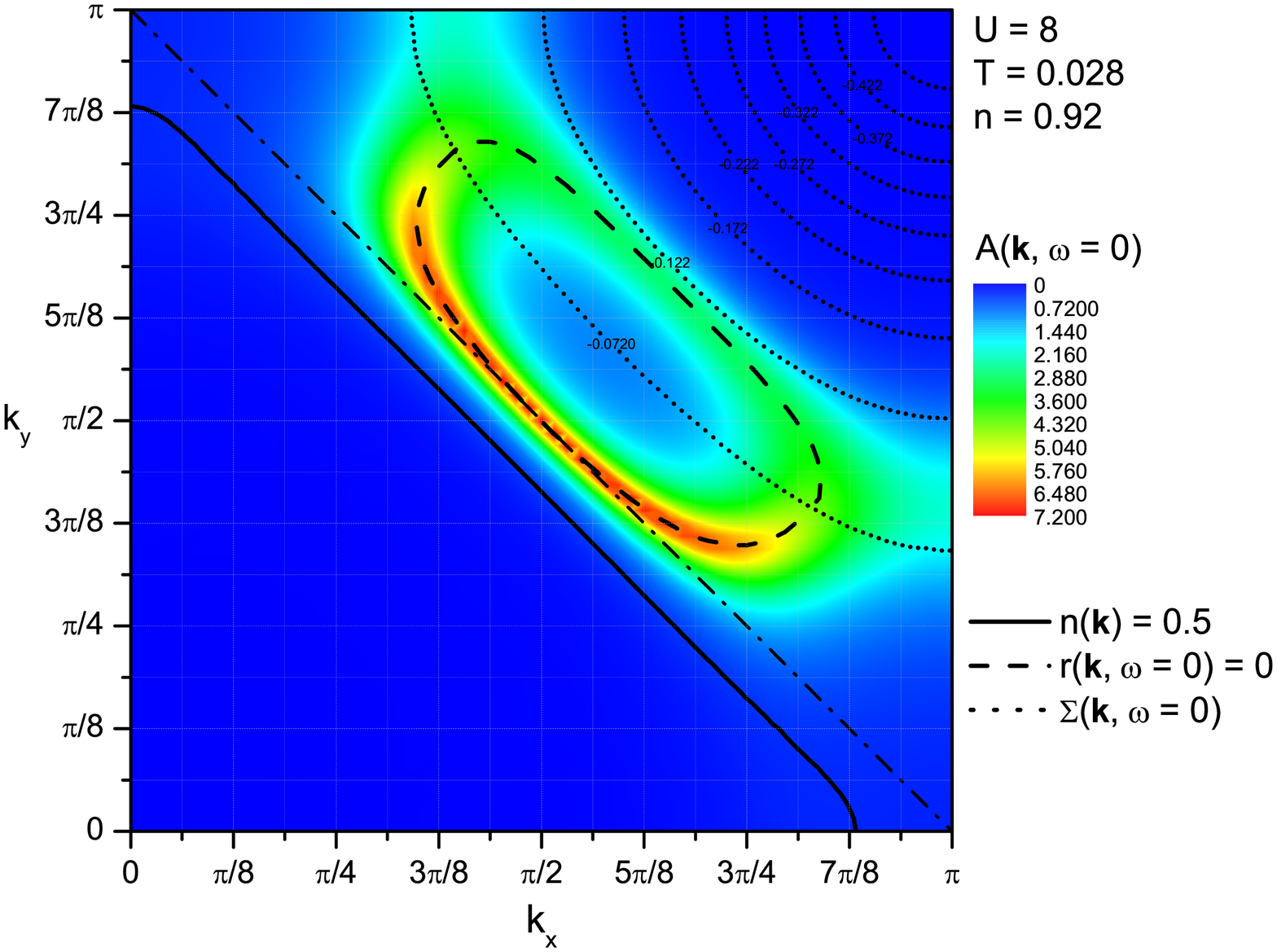}
\caption{Spectral function at the chemical potential
$A(\mathbf{k},\omega=0)$ as a function of momentum $\mathbf{k}$ for
$U=8$, (left) $n=0.78$  and $T=0.01$ (center) $n=0.85$  and $T=0.01$
(right) $n=0.92$ and $T=0.028$.}\label{fig2}
\end{figure}

The propagator $G$ is computed through the self-consistency scheme
depicted in Fig.~\ref{fig1}: we first compute $G_0$ and $B_0$ in
two-pole approximation, then $\Sigma$ and consequently $G$. Finally,
we check how much the fermionic parameters ($\mu$, $\Delta$ and $p$)
changed and decide if to stop or to continue by computing new $B$
and $\Sigma$ after $G$ and so on. To get $6$-digit precision for
fermionic parameters, we usually need about $10$ cycles (it varies
very much with doping, temperature and interaction strength) on a 3D
grid of $128 \times 128$ points in momentum space and $4096$
Matsubara frequencies.

In Fig.~\ref{fig2}, we report the electronic spectral function at
the chemical potential $A(\mathbf{k},\omega=0)=-\frac{1}{\pi} \Im
\left[G_{cc}(\mathbf{k},\omega=0)\right]$ as a function of momentum
$\mathbf{k}$ for $U=8$, $n=0.78$ and $T=0.01$ (right panel),
$n=0.85$ and $T=0.01$ (middle panel) and $n=0.92$ and $T=0.02$ (left
panel). $G_{cc}=G_{11}+G_{12}+G_{21}+G_{22}$ is the electronic
Green's function. The maxima of $A(\mathbf{k},\omega=0)$ mark the
effective Fermi surface as measured by ARPES. The solid line marks
the level $0.5$ of the electronic momentum distribution function
$n(\mathbf{k})$ per spin, that is, the Fermi surface in a perfect
Fermi liquid. The dashed line marks the level zero of
$r(\mathbf{k})=\varepsilon_0(\mathbf{k})+\Sigma'_{cc}(\mathbf{k},\omega=0)$,
that is, the Fermi surface if no damping would be present. The
dotted lines are labeled with the values of
$\Sigma''_{cc}(\mathbf{k},\omega=0)$. The dashed-dotted line is a
guide to the eye and marks the reduced (antiferromagnetic) Brillouin
zone. $\varepsilon_0(\mathbf{k})=-4t\alpha(\mathbf{k})-\mu$ is the
noninteracting dispersion. $\Sigma_{cc}(\mathbf{k},\omega)$ is the
electronic self-energy
\begin{equation}
G_{cc}(\mathbf{k},\omega)=\left(\omega-\varepsilon_0(\mathbf{k})-\Sigma_{cc}(\mathbf{k},\omega)\right)^{-1}
\end{equation}

At large doping ($n=0.78$), we identify a weakly-interacting Fermi
metal. The Fermi surface, that marked by maxima of
$A(\mathbf{k},\omega=0)$, is practically coincident with the level
$0.5$ of the momentum distribution function. The rather low signal
in proximity of $M=(\pi,\,\pi)$ is reminiscent of the shadow band
(see Fig.~\ref{fig5}). At $n=0.85$, we just passed through optimal
doping ($n \cong 0.82$). This latter is marked by a change in the
topology of the Fermi surface between open and close and,
consequently, by the coincidence between the value of the chemical
potential and the position of the van Hove singularity (see
Fig.~\ref{fig5}). The chemical potential presents an inflection
point at this doping (not shown) which allowed us to determine its
position with accuracy. A certain discrepancy between the Fermi
surface and the level $0.5$ of the momentum distribution function is
now clearly visible around the antinodal points ($X=(\pi,\,0)$ and
$Y=(0,\,\pi)$). At low doping ($n=0.92$), the situation is
dramatically changed and the scenario is that of a
strongly-interacting antiferromagnetic metal. The Fermi surface is
ill defined (it does not enclose a definite region of momentum
space) and does not coincide with the level $0.5$ of the momentum
distribution function: we have no more a Fermi liquid. The formation
of a pseudogap can be deduced by the remarkable difference between
the intensities at the cold spots (the well defined arch departing
from $(\pi/2,\pi/2)$, the nodal point) and the hot spots (the
regions in proximity of the antinodal points). The imaginary part of
the self-energy is so intense on the outer part of the hole pocket
(the Fermi surface if no damping would be present) to reduce it just
to an arch as reported by ARPES experiments \cite{Damascelli_03}.
The antiferromagnetic fluctuations are so strong to destroy the
coherence of the quasi-particles in that region of momentum space.

\begin{figure}[tb]
\begin{center}
\includegraphics[width=\columnwidth]{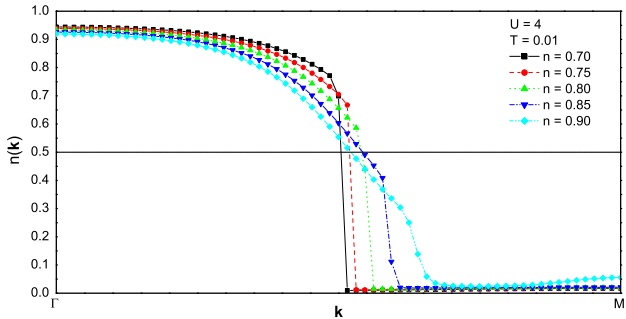}
\end{center}
\caption{Momentum distribution function $n(\mathbf{k})$ along the
principal diagonal of the Brillouin zone for various fillings at
$U=4$ and $T=0.01$.}\label{fig3}
\end{figure}

In Fig.~\ref{fig3}, we report the electronic momentum distribution
function $n(\mathbf{k})$ per spin as a function of momentum
$\mathbf{k}$ along the principal diagonal of the Brillouin zone
$\Gamma=(0,\,0) \to M=(\pi,\,\pi)$ for various fillings at $U=4$ and
$T=0.01$. On increasing the filling (reducing the doping) the quite
sharp jump going through the level $0.5$, clearly visible for
$n=0.7$, progressively moves its center downward and almost
disappears for $n=0.9$. In particular, between $n=0.8$ and $n=0.85$,
we can clearly see the appearance of a finite slope at the level
$0.5$, signalling the passage from Fermi-liquid excitations to
non-Fermi-liquid ones. The Fermi surface itself becomes ill defined.
The formation of a hole pocket for the lowest doping is signalled by
the appearance of finite weight at $M=(\pi,\,\pi)$.

\begin{figure}[tb]
\begin{center}
\includegraphics[width=\columnwidth]{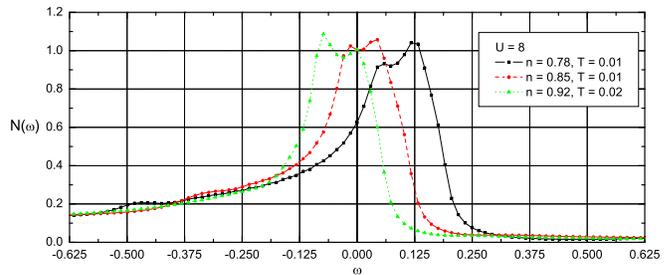}
\end{center}
\caption{Density of states for $U=8$, (solid line) $n=0.78$  and
$T=0.01$ (dashed line) $n=0.85$  and $T=0.01$ (dotted line) $n=0.92$
and $T=0.02$.}\label{fig4}
\end{figure}

In Fig.~\ref{fig4}, we report the electronic density of states
$N(\omega)$ per spin as a function of frequency for $U=8$, (solid
line) $n=0.78$ and $T=0.01$ (dashed line) $n=0.85$  and $T=0.01$
(dotted line) $n=0.92$ and $T=0.02$ in the frequency region in
proximity of the chemical potential. On increasing the filling
(reducing the doping) there is an evident transfer of spectral
weight between the top of the dispersion band (see Fig.~\ref{fig5})
and the antinodal points where the van Hove singularity resides. At
the lowest doping ($n=0.92$), we clearly see a well developed
pseudogap below the chemical potential as a depression between two
peaks, one pinned at the Fermi surface and another at the van Hove
singularity.

\begin{figure}[tb]
\includegraphics[width=\columnwidth]{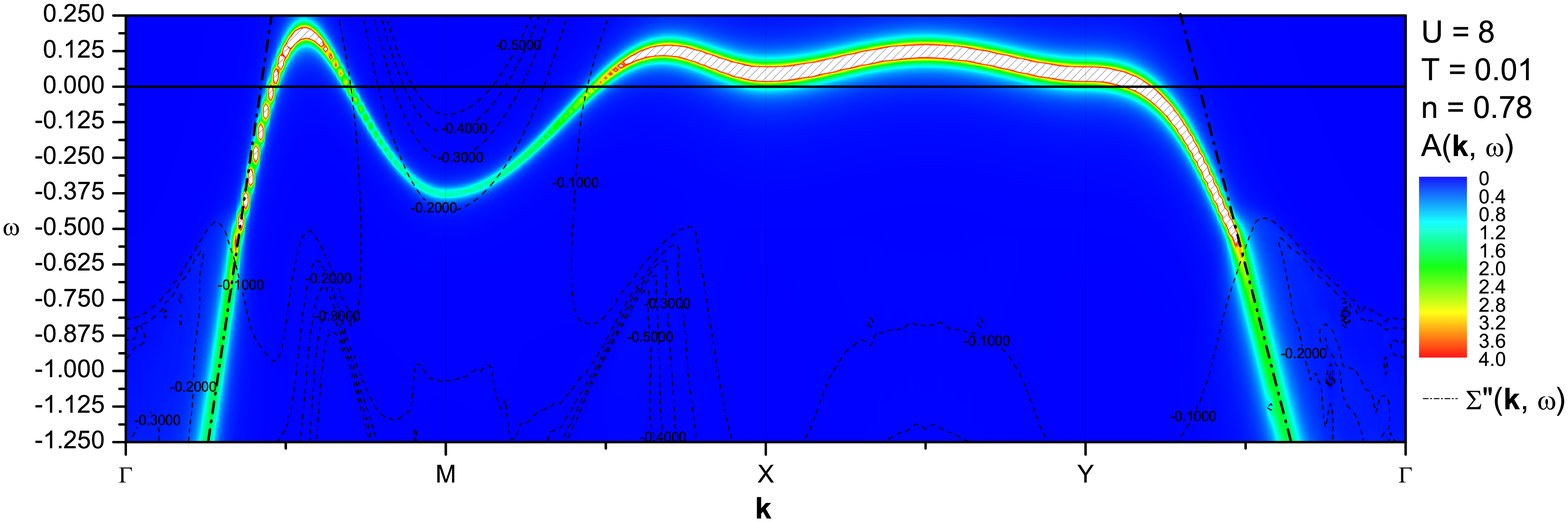}
\includegraphics[width=\columnwidth]{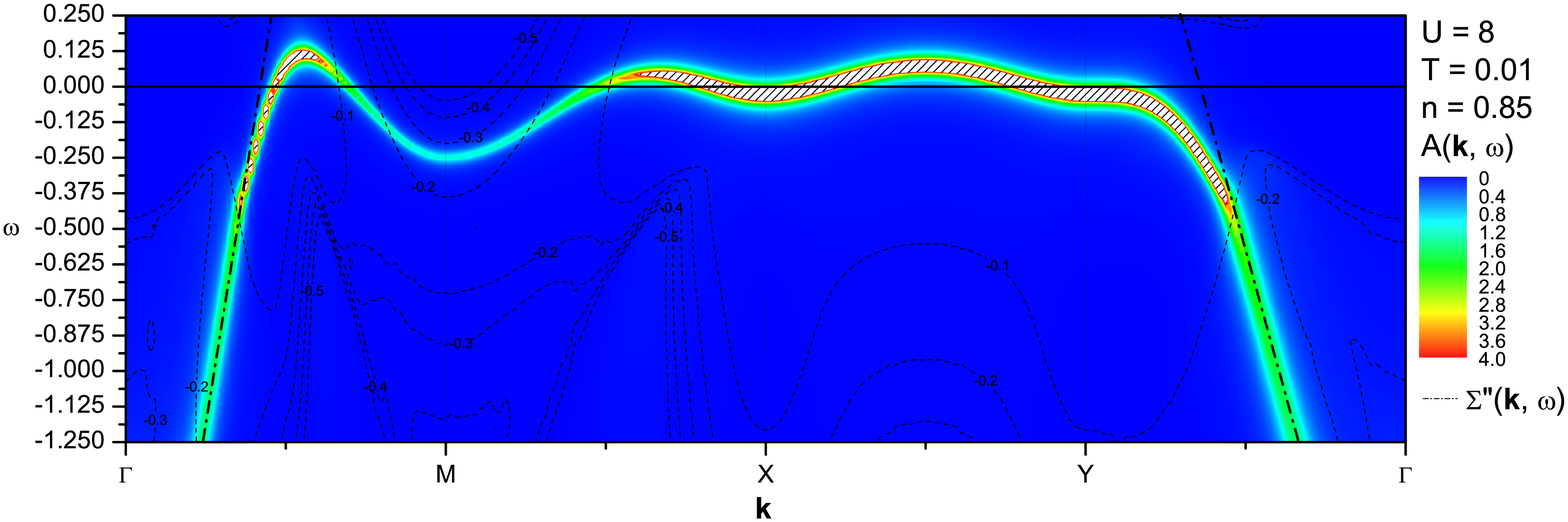}
\includegraphics[width=\columnwidth]{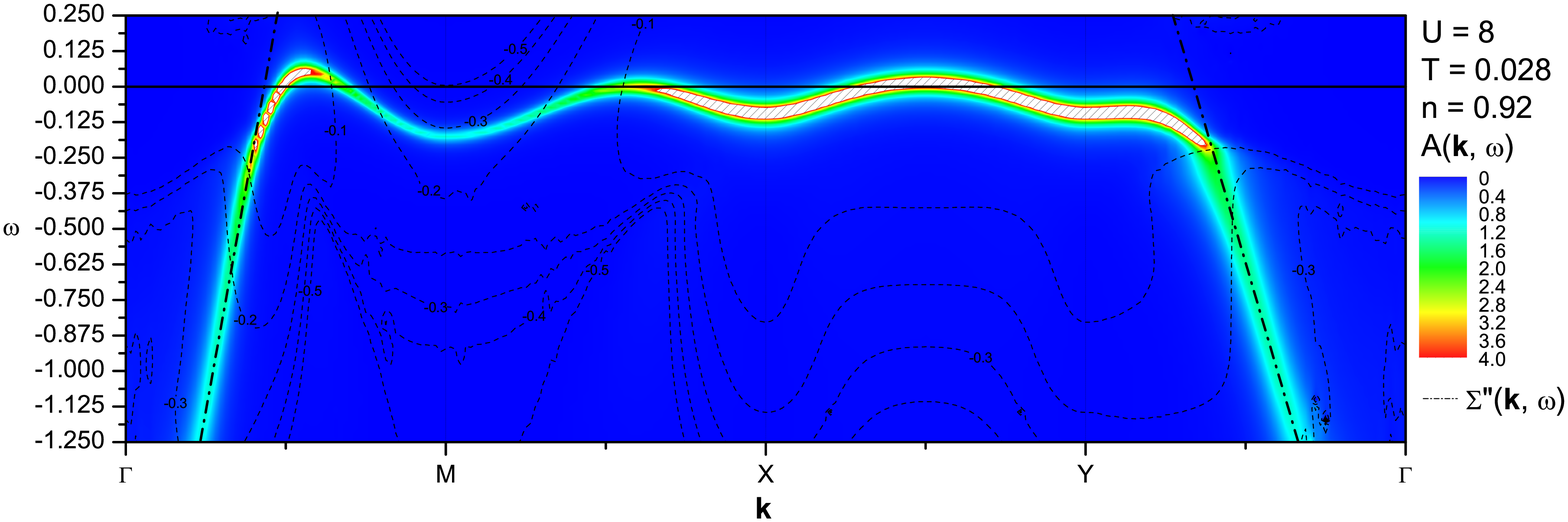}
\caption{Spectral function $A(\mathbf{k},\omega)$ along principal
directions for $U=8$, (left) $n=0.78$  and $T=0.01$ (center)
$n=0.85$  and $T=0.01$ (right) $n=0.92$ and $T=0.028$.}\label{fig5}
\end{figure}

In Fig.~\ref{fig5}, we report the electronic spectral function
$A(\mathbf{k},\omega)$ along the principal directions
($\Gamma=(0,\,0) \to M=(\pi,\,\pi)$, $M=(\pi,\,\pi) \to
X=(\pi,\,0)$, $X=(\pi,\,0) \to Y=(0,\,\pi)$ and $Y=(0,\,\pi) \to
\Gamma=(0,\,0)$) for $U=8$, $n=0.78$ and $T=0.01$ (top panel),
$n=0.85$ and $T=0.01$ (middle panel) and $n=0.92$ and $T=0.02$
(bottom panel) in proximity of the chemical potential. The dashed
lines are labeled with the values of $\Sigma''(\mathbf{k},\omega)$.
The dashed-dotted line is a guide to the eye and marks the direction
of the dispersion just \emph{before} the kink. The presence of kinks
in the dispersion in both the nodal ($\Gamma=(0,\,0) \to
M=(\pi,\,\pi)$) and the antinodal ($X=(\pi,\,0) \to \Gamma=(0,\,0)$)
directions is quite evident and signals the coupling of the
electrons to a bosonic mode as reported by ARPES experiments
\cite{Damascelli_03}. In our formulation, the mode is clearly
magnetic. The extension of the flat region in the dispersion around
the antinodal points increases systematically on decreasing doping.
This clearly signals the transfer of spectral weight from the Fermi
surface as it is destroyed by strong correlations (see
Fig.~\ref{fig4}).

In conclusion, we have shown how a pseudogap scenario and
non-Fermi-liquid features can be obtained in the 2D Hubbard model
within the Composite Operator Method with the electronic self-energy
computed in the Self-Consistent Born Approximation. This scenario is
just the one recently claimed for underdoped Cuprates by ARPES
experiments \cite{Damascelli_03}. In particular, we report:
formation of a pseudogap with related \emph{hot} an \emph{cold}
spots and \emph{arcs} on the Fermi surface; non-Fermi liquid
features such as the non coincidence of the level $0.5$ of the
momentum distribution function and the effective Fermi surface and
as the absence of a jump in the momentum distribution function at
the level $0.5$; kinks in the dispersion along nodal and anti-nodal
directions. We are now planning to compute the residual self-energy
of the bosonic propagators and to take into account the
next-nearest-neighbor hopping term in the Hamiltonian in order to
make quantitative comparisons with experiments.

\begin{acknowledgments}
We wish to gratefully acknowledge many useful discussions with N. M.
Plakida and P. Prelovseck.
\end{acknowledgments}

\bibliographystyle{apsrev}
\bibliography{biblio}

\end{document}